\documentclass[a4paper,fleqn,usenatbib]{mnras}

\usepackage[T1]{fontenc}
\usepackage{ae,aecompl}
\usepackage{ulem,soul}

\usepackage{graphicx}	
\usepackage{amsmath}	
\usepackage{amssymb}	

\usepackage{graphicx,color}
\usepackage[usenames,dvipsnames,svgnames,table]{xcolor}

\title[Structure of PPDs with winds]{Structure of Protoplanetary Discs with Magnetically-driven  Winds}
\author[F. Khajenabi et al.]{
Fazeleh Khajenabi$^{1,2}$,\thanks{E-mail: f.khajenabi@gu.ac.ir}
Mohsen Shadmehri$^1$,
Martin E. Pessah$^{3}$, 
Rebecca G. Martin$^{4}$
\\
$^1$Department of Physics, Faculty of Sciences, Golestan University, Gorgan 49138-15739, Iran\\
$^2$Research Institute for Astronomy and Astrophysics of Maragha (RIAAM), Maragha, P.O. Box: 55134-441, Iran\\
$^{3}$Niels Bohr International Academy, Niels Bohr Institute, Blegdamsvej 17, DK-2100, Copenhagen~\O, Denmark\\
$^{4}$Department of Physics and Astronomy, University of Nevada, Las Vegas, 4505 South Maryland Parkway, Las Vegas, NV 89154, USA\\
}

\date{Accepted XXX. Received YYY; in original form ZZZ}

\pubyear{2017}

\begin{document}
\label{firstpage}
\pagerange{\pageref{firstpage}--\pageref{lastpage}}
\maketitle

\begin{abstract}
We present a new set of analytical solutions to model the steady state structure of a protoplanetary disc with a magnetically-driven wind. Our model implements a parametrization of the stresses involved and the wind launching mechanism in terms of the plasma parameter at the disc midplane, as suggested by the results of recent, local MHD simulations. When wind mass-loss is accounted for, we find that its rate significantly reduces the disc surface density, particularly in the inner disc region.  We also find that models that include wind mass-loss lead to thinner dust layers. As an astrophysical application of our models, we address the case of HL Tau, whose disc exhibits a high accretion rate and efficient dust settling at its midplane. These two observational features are not easy to reconcile with conventional accretion disc theory, where the level of turbulence needed to explain the high accretion rate would prevent a thin dust layer. Our disc model that incorporates both mass-loss and angular momentum removal by a wind is able to account for HL Tau observational constraints concerning its high accretion rate and dust layer thinness. 

\end{abstract}




\section{Introduction}
Our ability to characterize protoplanetary discs (PPDs) is undergoing a revolution with the advent of the Atacama Large Millimeter Array (ALMA). Understanding the structure of these discs is critical in order to shed light into the processes that shape the formation and evolution of planetary systems, including our own Solar System.
%
An essential part of any model for describing the structure and evolution of an accretion disc is the mechanism that enables the transport of angular momentum.
Nearly three decades since the magnetorotational instability \citep[MRI;][]{Balbus91} was introduced  as the dominant mechanism for angular momentum transport in accretion discs, and despite of its significant success, it turns out that under certain conditions like insufficient level of ionization, this mechanism loses its effectiveness \citep[see, e.g.,][]{Gammie96,Armitage01, 2007MNRAS.378.1471L,
2008ApJ...684..498P, Martin12, 2014A&A...566A..56L, 2017ApJ...838...48M}. Thus, other angular momentum transport mechanisms may be needed for efficient gas accretion. For instance, the outer parts of a PPD are gravitationally unstable, and so, gravitational instability may play a dominant role in transporting angular momentum  \citep[e.g.,][]{Lodato2004,Rice2005,Rafikov2005,Cai2010,Martin14,Rice2014,Rafikov15,Dong16}. 

While  there are still many uncertainties about the true nature of the processes driving mass accretion in PPDs, observational evidence suggests that there are winds or outflows in these systems that seem to be related to the accretion process itself \citep[e.g.,][]{Matsakos16,Wang17}. 
Wind models are generally categorized according to their launching mechanism and as referred to as radiation-driven and magnetically-driven outflows, where the former primarily being important in the outer region of a PPD and possibly in the final stages of its evolution, and the latter is effective in the inner region of a PPD.
A wind emanating from   an accretion disc, regardless of its launching mechanism, is able to extract part of the accreting mass, disc angular momentum, and even convert energy released in the accretion process into kinetic energy. The structure and evolution of the underlying disc are affected significantly by the wind \citep[e.g.,][]{Lubow94b,Combet2008,Salmeron10}. 

Understanding the connection between the processes that enable efficient accretion and those responsible for the launching of the wind remains an active research area. Because constructing a unified theory that simultaneously describes these two phenomena at a global scale still remains in the horizon, it is not unusual for theorists to investigate the structure of a disc and its wind as separate, but related, processes.
The relationship between these is usually described in a phenomenological way based on physical considerations or models which are valid under restricted conditions \citep[e.g.,][]{Blandford82,Konigl89,Pelletier92,Knigge99,Ogilvie2001,Combet2008,Salmeron10,Lesur13,Bai-etal2016}. Numerical simulations provide a useful complementary tool to investigate the intertwined processes governing discs with winds \citep[e.g.,][]{Proga98,Suzuki2010,Simon13a, Simon13, Bai-Stone13,Lesur2017}.

The capability of the conventional disc theory to explain the observationally expected rapid accretion in PPDs has been challenged by recent simulations that have taken into account non--ideal magnetohydrodynamical (MHD) effects \citep[e.g.,][]{Bai-Stone13,Bai13,Lesur2017}. With shearing-box MHD simulations of PPDs including Ohmic resistivity and ambipolar diffusion, \cite{Bai-Stone13} showed that not only is the MRI suppressed but also a strong magnetocentrifugal wind is launched and the accretion is driven mostly by the wind. It then turns out that wind-driven accretion is primarily efficient in the inner parts of a PPD, whereas farther away from the central star a combination of the MRI and the disc wind is responsible for the gas accretion \citep{Bai13}. Subsequent  non--ideal MHD simulations showed that the configuration of the large scale magnetic field has a vital role in driving or suppressing the accretion \citep{Lesur2017}. We note that the idea of wind-driven accretion has not been suggested so recently and has already been raised by a few authors  \citep[e.g.,][]{Lubow94a, Salmeron07,Salmeron2007b,Lesur13}. However, the novelty is that recent numerical simulations of PPDs have provided a potential link between the wind--driven accretion paradigm  and the non--ideal MHD effects. Efforts to describe observational features of  PPDs using the wind-driven accretion scenario  are gradually presenting their preliminary results. For instance, \cite{Wang17} showed that the inner cavities in transitional PPDs are the natural outcome of the wind-driven accretion theory.

Although numerical simulations are promising in describing a PPD structure and wind launching, they also have their own limitations that make it difficult to have a direct comparison of the results with the observations. In addition, most of the current simulations of PPDs are usually done using the shearing--box approximation, which can not capture global phenomena. For these reasons, and given the simplicity of constructing a standard accretion disc model, many authors  try to present models for PPDs within the framework of the standard disc model. However, the free parameters of their models, which are introduced due to lack of detailed knowledge about the disc turbulence and its link to potential wind launching mechanisms, are generally constrained by the results of the   existing MHD simulations of the PPDs.

Based on this new theoretical framework,  \cite{Suzuki16} investigated the evolution of a PPD with  magnetically-driven winds and heating due to turbulent viscosity. They also parameterized the wind mass-loss and angular momentum transport by the disc turbulence and the wind. Although these processes are not independent, \cite{Suzuki16} adopted values of these input parameters independently, but based on the local disc simulations they found a large range of surface density profiles. For certain sets of the input parameters, the surface density profile exhibits a positive slope due to presence of a wind. The astrophysical implications regarding dust dynamics and planet migration are worthwhile to be studied further. For instance,  \cite{Ogihara2015} showed that type~I migration of a planetary embryo becomes slower or even halts  in an evolving disc with the wind \citep[also see,][]{Ogihara2015b}. However, they did not focus on exploring the dust dynamics in a disc with wind.  \cite{Pinilla16} investigated the dust dynamics in a PPD with a dead zone and magnetically-driven winds. Their model has also been constructed within the framework of the  standard disc theory, however, they showed that gaps and asymmetries in the  transitional discs can be explained as a consequence  of considering a dead zone and a magnetically-driven wind. They did not consider angular momentum transport by the wind and the input parameters related to the disc turbulence and wind mass-loss treated independently. 

Magnetic wind-driven accretion disc models have also been proposed to explain certain observational features of HL Tau \citep{Matsakos16,Hasegawa17}. The accretion rate on to the central star is observed to be around $8.7 \times 10^{-8}\,\rm M_\odot\, yr^{-1}$ \citep{Beck2010}. Radiative transfer modelling of ALMA data suggests that the scale height of millimeter sized grains is only about $1\,\rm AU$ at a radial distance of $100\,\rm AU$ \citep{Pinte16}. For significant  settling to occur, turbulence in the disc must not be strong.  \cite{Hasegawa17} (hereafter H17) argue that the current high degree of dust settling in HL Tau  can not be compatible with the estimated high accretion rate onto the central star in the framework of the conventional viscous disc model, unless a magnetically-driven wind plays  a significant role in transporting disc angular momentum. The spirit of their analysis is similar to that of \cite{Armitage13}, in which angular momentum transport by both the disc turbulence and winds are permitted. In light of the recent MHD disc simulations, these key transport mechanisms are parameterized in terms of the plasma parameter at the disc midplane. The model of \cite{Armitage13} follows the disc evolution in the presence of  magnetically-driven winds, however, the analysis of H17 constructed an approximate steady state model. We extend these works to include wind mass-loss in to the model. 

Our goal is to explore the steady state structure of a PPD in the presence of a magnetically-driven wind. Our basic assumptions are similar to H17 and \cite{Armitage13}, however, our models also account for wind mass loss. The rest of the paper is organized as follows.
In Section~2, we present our main assumptions and equations defining the disc model with winds.
In Section 3, we find analytical solutions for the steady-state disc-surface density both in the case where wind mass loss is negligible and when it is efficient. 
In Section 4, we discuss the astrophysical implications of our model. We provide our concluding remarks in Section~5.

\section{General Formulation}
Our disc model is based on standard accretion disc theory. Our approach, however, goes beyond the standard $\alpha$-disk model where the efficiency of angular momentum transport is encapsulated in a constant parameter. Based on the results of shearing-box simulations, we implement a parametrization of the stresses involved and the wind launching in terms of the plasma parameter $\beta_0$ at the disc midplane \citep{Fromang13}. The plasma parameter, $\beta_0$, is defined as the ratio of gas to the magnetic pressure at the disc midplane. Following recent MHD disc simulations \citep{Lesur2017}, we furthermore include a wind mass-loss rate that depends on $\beta_0$.

\cite{Suzuki16} showed that the equation governing the evolution of the disc surface density may be written as 
\begin{align}
\frac{\partial\Sigma}{\partial t}- & \frac{1}{r}\frac{\partial}{\partial r} \left [ \frac{2}{r\Omega_{\rm K}}  \frac{\partial}{\partial r} (r^2 \Sigma c_{\rm s}^{2} W_{ r\phi})  + \frac{2r}{\Omega_{\rm K}} (\rho c_{\rm s}^2)_{\rm mid} W_{ z\phi} \right] \notag \\ 
& + C_{\rm w} (\rho c_{\rm s})_{\rm mid} =0
\label{eq:mainI}
\end{align}
\citep[see also,][]{Suzuki2010,Fromang13}, where $W_{r\phi}$ and $W_{z\phi}$ are the normalized accretion stresses of the disc MHD turbulence and the wind launching, respectively. These stress components are defined in terms of time-averaged disc quantities \citep[][]{Fromang13,Suzuki16}. Furthermore, a dimensionless parameter $C_{\rm w}$ is introduced to quantify the rate of wind mass-loss.  Here, the gas volume density, surface density and the sound speed are denoted by $\rho$, $\Sigma$ and $c_{\rm s}$, respectively. Quantities with index "mid" are evaluated at the disc midplane. In particular, the midplane density is $\rho_{\rm mid}=\Sigma/(\sqrt[]{\pi} H)$, where $H$ is disc thickness given by $H=\sqrt{2}c_{\rm s}/\Omega_{\rm K}$. 

Regarding the rotation profile of the disc, we assume that the angular velocity is Keplerian for all radii, i.e., $\Omega_{\rm K}=\sqrt{GM_{\star}/r^3}$, where $r$ is the radial, cylindrical distance from the central star and $M_{\star}$ is the mass of the star.
This is a reasonable approximation within the bulk of the disc, where the gas pressure gradient is much smaller than the gravitational force exerted by the central star. We note that the rotational velocity may deviate from a Keplerian profile if the disc self-gravity or magnetic forces play an important dynamical role or if the radial gas pressure gradient becomes significant. In the present study we neglect disc self-gravity and the magnetic forces that would be associated with the stresses involved are very weak for the plasma $\beta_0$ we considered. 
%


For simplicity, we do not consider the energy balance of the disc, and adopt instead a prescribed power-law radial temperature profile so that
\begin{equation}
T=T_0 \left (\frac{r}{r_0} \right )^{-q},
\end{equation} 
where we take $T_0 = 100$ K and $r_0 = 1$ AU. 
Although we present disc solutions for an arbitrary temperature exponent $q$, our focus is on two given values for this exponent which have been used widely in the previous studies: $q=1/2$ and $q=1$. The case with $q=1/2$  occurs when viscous heating is negligible compared to the stellar irradiation \citep[e.g.,][]{Frank}. The sound speed may then be written as $c_{\rm s}=(k_{\rm B} T/\mu m_{\rm H})^{1/2}$, where $k_{\rm B}$ and $m_{\rm H}$ are the Boltzmann constant and the mass of the hydrogen atom, respectively. Since most of the mass of the disc is in molecular hydrogen, the mean molecular weight is $\mu=2.3$ \citep[e.g.,][]{Armi}.

In steady state, equation (\ref{eq:mainI}) can be integrated in radius to obtain
\begin{equation}\label{Mdot_total}
\dot{M}_{\rm d}(r) + \dot{M}_{\rm w}(r) = \textrm{constant} = \dot{M}_{\rm d}(r_{\rm out}),
\end{equation}
where $\dot M_{\rm d}(r)$ is the mass accretion rate through the disc and $\dot M_{\rm w}(r)$ is the mass loss rate leaving the disc in the wind, that both depend upon the radius, $r$.   The constant accretion inflow rate at the outer edge of the disc is denoted by $\dot M_{\rm d}(r_{\rm out})$. Note that in order for the steady state to be a good approximation, it is required that any changes in the mass accretion rate at the outer boundary take place on timescales that are much longer than the timescale required for the global disk structure to relax to a new steady state.

The mass accretion rate through the disc $\dot{M}_{\rm d}(r)$ has two contributions, \citep[e.g.,][]{Suzuki16},
\begin{equation}\label{Mdot}
\dot{M}_{\rm d}(r)=\frac{4\pi}{r\Omega_{\rm K}} \frac{\partial}{\partial r} (r^2 \Sigma c_{\rm s} W_{r\phi} ) +\frac{4\pi r}{\Omega_{\rm K}} (\rho c_{\rm s}^2)_{\rm mid} W_{z\phi}.
\end{equation}
The first term corresponds to the mass accretion rate driven by disc turbulence, and the second term is the contribution of the wind stress to the total mass accretion rate. 
The term $\dot{M}_{\rm w}(r)$ corresponds to the mass-loss rate by the wind,
\begin{equation}\label{Mdot_z}
\dot{M}_{\rm w}(r) = 2\pi \int_r^{r_{\rm out}} r\, C_{\rm w} (\rho c_{\rm s})_{\rm mid} \, dr.
\end{equation}
The key assumption that goes into writing equation $(\ref{Mdot_total})$ is that $\dot{M}_{\rm w}(r)$ vanishes at $r=r_{\rm out}$, i.e., $\dot{M}_{\rm w}(r_{\rm out}) = 0$ and $\dot{M}_{\rm d}(r)$ goes to a well-defined value $\dot{M}_{\rm d}(r_{\rm out})$. Note that under this assumption, $(\ref{Mdot_total})$ is a natural way to write the outer boundary condition whether or not there is a wind.

%
%

%

\begin{figure}
\includegraphics[scale=0.55]{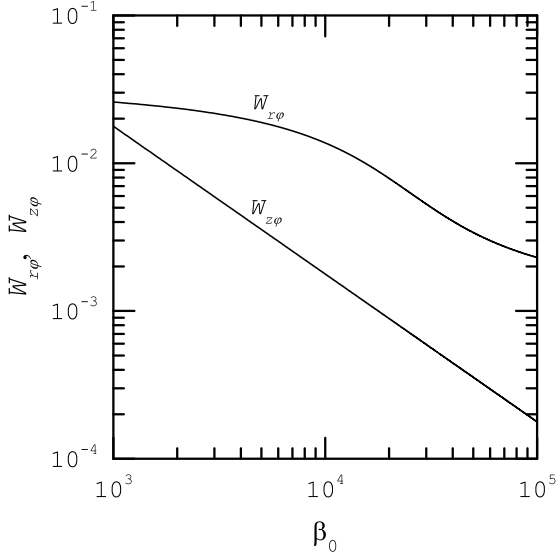}
\caption{Stress components $W_{r\phi}$ and $W_{z\phi}$ as a function of $\beta_0$, as 
described by equations (\ref{eq:wrphi}) and (\ref{eq:wzphi}).
}\label{fig:ww}
\end{figure}

According to MHD shearing-box disc simulations, the stresses associated with disc turbulence, $W_{r\phi}$, and wind launching, $W_{ z\phi}$, can be parameterized as
%
\begin{equation}
\log W_{r\phi} = -2.2 + 0.5 \tan^{-1} \left(\frac{4.4-\log\beta_0}{0.5}\right),
\label{eq:wrphi}
\end{equation}
and
\begin{equation}
\log W_{ z\phi}=1.25-\log\beta_0 ,
\label{eq:wzphi}
\end{equation}
\citep{Armitage13}, where the plasma parameter is written as $\beta_0  = 8\pi P_{\rm mid}/B_{\rm z}^2$. Here, $P_{\rm mid}$ is the gas pressure at the disc midplane and $B_{\rm z}$ is the vertical component of the magnetic field.  Fig. \ref{fig:ww} shows the profiles for the stress components $W_{r\phi}$ and $W_{z\phi}$ as a function of $\beta_0$. The dominant contribution to the stress is due to disc turbulence for large values of $\beta_0$. As $\beta_0$ decreases, the fractional contribution to the stress due to the disc wind becomes more significant.

\cite{Armitage13} used these relations to investigate the evolution of a time--dependent disc within the framework of the standard accretion theory neglecting wind mass-loss. The inclusion of the wind mass-loss enables us to explore how the steady-state disc structure is modified when a fraction of the accreted mass is lost by the wind. Although numerical simulations suggest that the wind mass-loss rate depends primarily on the average ratio of magnetic to thermal pressure in the disk midplane \citep{Lesur2017}, there is currently no parameterization available to account for this effect. As a simple approximation, therefore, we assume that $C_{\rm w}$ is proportional to the stress due to the wind,  where the constant of proportionality is an input parameter of the model.

We transform equation (\ref{eq:mainI}) into a dimensionless form with the dimensionless variables
\begin{equation}\label{eq:reference}
y=\frac{\Sigma}{\Sigma_0}, ~~x=\frac{r}{r_0}, ~~ m_\star=\frac{M_{\star}}{M_{\odot}}, ~~{\rm and}~~ \tau=\frac{t}{t_0},
\end{equation}
where we adopt $\Sigma_0 = 1700$ g cm$^{-2}$ and $t_0 = 1/\Omega_0$. The Keplerian angular velocity at $r_0$ is denoted by $\Omega_0$.

Using the above definitions, we can obtain dimensionless expressions for equation (\ref{eq:mainI}), governing the evolution of the disc surface density,
\begin{align}
\frac{\partial y}{\partial\tau} - 
\frac{1}{x} \frac{\partial}{\partial x} \left [ \sqrt{x} \frac{\partial}{\partial x} \left ( \xi_{ r} x^{2-q} y \right )+ \xi_{ z} x^{(2-q)/2} y \right ] 
+ \xi_{ w} x^{-3/2} y=0, 
\label{eq:main}
\end{align}
and equation (\ref{Mdot}), providing the scaled mass accretion rate through the disc,
\begin{equation}\label{eq:accretion-rate}
\dot{m}_{\rm d}(x)=\sqrt{x} \frac{\partial}{\partial x} \left (\xi_{r} x^{2-q} y \right ) + \xi_z x^{(2-q)/2} y ,
\end{equation}
where $\dot{m}_{\rm d}(x)=\dot{M}_{\rm d}(r)/(2\pi \Sigma_{0} r_{0}^2 \Omega_{0})$ is a function of the radial distance.

To simplify the equations we introduce the dimensionless parameters 
\begin{equation}
\xi_{ r}= \frac{1}{\sqrt{m_\star}} \left (\frac{H_0}{r_0} \right )^2 W_{ r\phi} (\beta_0 ),
\end{equation}
\begin{equation}
\xi_{ z}= \frac{1}{\sqrt{\pi}} \left (\frac{H_0}{r_0} \right ) W_{ z\phi} (\beta_0 ),
\end{equation}
and
\begin{equation}
\xi_{ w}=\sqrt{\frac{m_\star}{2\pi}} C_{\rm w}.
\end{equation}
Using these definitions, we can solve equation (\ref{eq:main}) as a function of the plasma parameter, $\beta_0$, subject to the appropriate boundary and/or initial conditions in the steady-state and time-dependent cases. In the next section, we present analytical solutions for a steady-state disc model with wind. We consider a disc that is in a steady state with a constant mass flux injected at its outer edge. If the wind mass-loss is negligible, this inward mass flow is all eventually accreted by the star. When the wind mass-loss is included, a fraction of the radial gas flow is accreted by the central star and the rest is ejected by the wind. This effect has important implications for the surface density profile of the underlying disk.

\section{Steady State Solutions}
In this section, we study analytically a steady state model for a disc with a magnetically-driven wind. Although the steady state approximation limits the applicability of our solutions, it is possible to use these solutions to describe certain observational features of a PPD. For instance, H17 used a steady-state model for a disc with a wind to explain the structure of HL Tau. In their approach, they considered a disc model with no wind mass-loss ($\xi_{\rm w}=0$) and employed a relatively crude approximation in order to solve equation (\ref{eq:main}) and obtain the density profile analytically. However, as we show below, their simplification is not necessary in order to obtain fully analytical solutions for a disc with a wind when the wind mass-loss is negligible.  We then generalize the study to a case with non-negligible wind mass-loss.  Our new analytical solutions demonstrate that the disc surface density presents a very different trend compared to a disc without mass-loss. 
%

In order to find steady-state solutions to our main equation (\ref{eq:mainI}), we need to specify two boundary conditions no matter whether $\xi_{w}$ is zero or non-zero. We fix the two constants of integration in the final solutions as follows. One of the constants is determined by requiring that the surface density vanishes at the inner boundary \citep{Pringle81}. In our dimensionless notation this is
\begin{equation}\label{eq:boundary_I}
y(x_{\rm in})=0,
\end{equation}
 where we assume that the inner radius is $x_{\rm in}=0.01$.
The other constant is specified by the accretion rate at a sufficiently large outer radius. 
This means that the gas is assumed to flow in at a constant rate. We note that when the wind mass-loss is included, the accretion rate $\dot{m}$ is no longer constant with radial distance. It is expected that in the presence of the wind, as we move toward the inner regions, the accretion rate gradually decreases. Thereby,  we can set the accretion at the outer edge, i.e. $\dot{m}(x_{\rm out})=\dot{m}_{\rm out}$, as a given input parameter. Equation (\ref{eq:accretion-rate}) then implies that
\begin{equation}
\label{outer-boundary-wind}
\left [ \sqrt{x} \frac{\partial}{\partial x} \left (\xi_{r} x^{2-q} y \right ) + \xi_z x^{(2-q)/2} y \right ]_{x_{\rm out}} = \dot{m}_{\rm out},
\end{equation}
where we set $x_{\rm out}=100$. The above equation is our second boundary condition.

\begin{figure}
\includegraphics[scale=0.65]{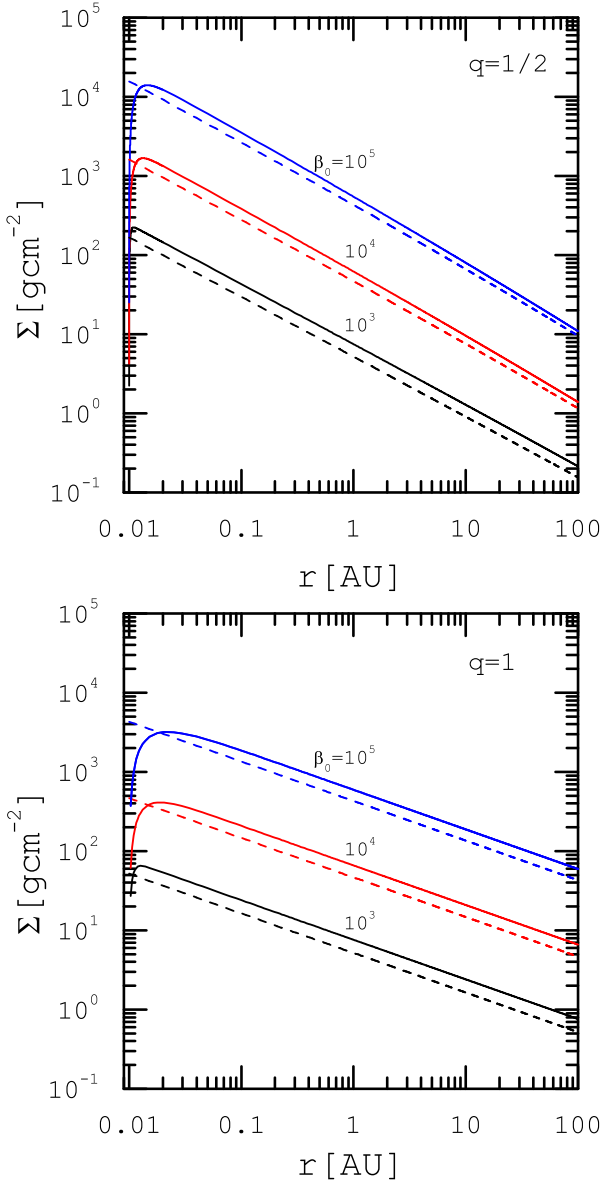}
\caption{Surface density profile as a function of the radial distance for $q=1/2$ (top panel) and $q=1$ (bottom panel). Solid curves represent exact surface density solutions for different values of $\beta_0$, whereas approximate H17 solutions  are shown by dashed lines for the same input parameters. The central mass is $M_{\star}=1$ M$_{\odot}$ and the accretion rate is assumed to be $\dot{M}_{\rm d}=10^{-8}$ M$_{\odot}$ yr$^{-1}$.}\label{fig:surf-a}
\end{figure}

\subsection{Solutions with no mass-loss, $\xi_{ w}=0$}
In the steady-state case with negligible wind mass loss, we set $\xi_{w}=0$ in equation (\ref{eq:main}). Thus, we obtain 
\begin{equation}\label{eq:steady-cw}
\sqrt{x} \frac{d}{dx} \left ( \xi_{ r} x^{2-q} y \right ) + \xi_{ z} x^{(2-q)/2} y = \dot{m}_{\rm d}(r_{\rm out})=\dot{m}_{\rm out},
\end{equation}
where the dimensionless accretion rate $\dot{m}_{\rm out}$ is a fixed given input parameter. This equation is analogous to equation (4) of H17, however, in the limit of $x\gg x_{\rm in}$,  they approximated  the first term of the left-hand side as $d( \xi_{ r} x^{2-q} y)/dx \approx (1/2)( \xi_{ r} x^{2-q} y)$. Then, the above equation reduces to an algebraic equation which can easily be solved to obtain the dimensionless surface density:
\begin{equation}
y(x)=2\,\dot{m}_{\rm out}\left [\xi_{r} x^{(5-2q)/2} + 2 \xi_z x^{(2-q)/2} \right ]^{-1}.
\end{equation}
\begin{figure}
\includegraphics[scale=0.6]{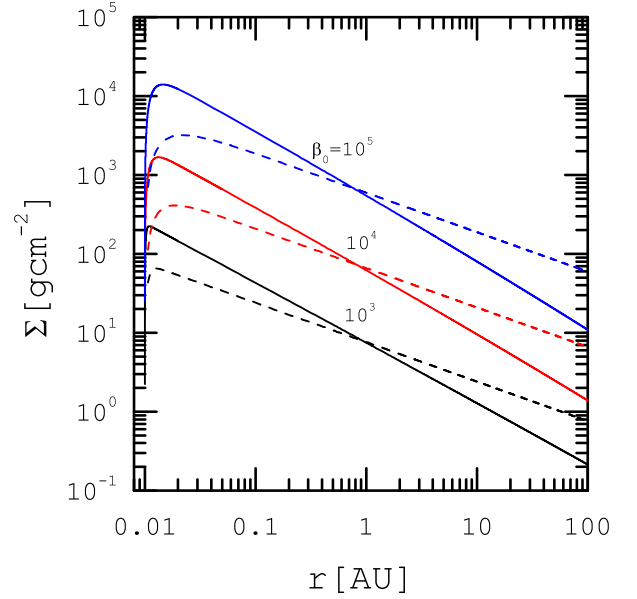}
\caption{Surface density as a function of the radial distance for $q=1/2$ (solid curves) and $q=1$ (dashed curves). Each curve is labeled by the corresponding value of $\beta_0$. The central mass is $M_{\star}=1$ M$_{\odot}$ and the accretion rate is assumed to be $\dot{M}_{\rm d}=10^{-8}$ M$_{\odot}$ yr$^{-1}$.
}
\label{fig:surf-b}
\end{figure}

Here, we do not use this approximation, rather we solve equation (\ref{eq:steady-cw}) as a first order differential equation. Fortunately, this  equation  is integrable and its general solution for $q\neq 1$ can be written as 
\begin{equation}\label{eq:sol-q}
y(x)= \frac{\dot{m}_{\rm out}\, x^{q-2}}{f(x)} \left ( \int \xi_{ r}^{-1} x^{-1/2}f(x) dx +C \right ),
\end{equation}
where the constant $C$ is obtained from the boundary condition, and the function $f(x)$ is defined as
\begin{equation}
f(x) =\exp \left[f_0 x^{(q-1)/2}\right],
\end{equation}
where $f_0 = 2 \xi_z / \xi_r (q-1)$.
When the temperature exponent is $q=1$, the general solution of equation (\ref{eq:steady-cw}) becomes
\begin{equation}
y(x) = \dot{m}_{\rm out} \left [ 2 (\xi_{ r} + 2\xi_{ z})^{-1} x^{-1/2} + C x^{-1-(\xi_{ z}/\xi_{ r})} \right ].
\end{equation}

Although the general solution (\ref{eq:sol-q}) is written for an arbitrary temperature exponent $q$, it is useful to simplify it for $q=1/2$ which is of particular interest. In this case, the solution becomes
%
%
\begin{multline}
y(x)=\dot{m}_{\rm out} 
\bigg \{ 
2 \xi_{ r}^{-1} x^{-1} ( 1 - 4 x^{-1/4} \xi_{ z}/\xi_{ r} ) 
%
+ \exp \left (4x^{1/4}\xi_{ z}/\xi_{ r} \right ) \times
\\
\left[32 \xi_{ z}^{2} \xi_{ r}^{-3} {\rm Ei}\left(1,4\xi_{ z}/\xi_{ r} x^{1/4}  \right) + C\right]  x^{-3/2} 
\bigg \},
\end{multline}
%
where ${\rm Ei}(1,4\xi_{ z}/\xi_{ r} x^{1/4} )$ is the exponential integral defined as 
\begin{equation}
{\rm Ei} (n,z)=\int_{1}^{\infty} t^{-n} \exp (-tz) dt. 
\end{equation}
The constant $C$ is obtained using the inner boundary condition (\ref{eq:boundary_I}).

Fig. \ref{fig:surf-a} displays the surface density profile versus radial distance for different temperature exponents  $q=1/2$ and 1. The central mass is $M_{\star}=1$ M$_{\odot}$ and the accretion rate is assumed to be $\dot{M}_{\rm d}=10^{-8}$ M$_{\odot}$ yr$^{-1}$. In the top panel, the surface density is shown by solid curves for $q=1/2$ and varying plasma parameter $\beta_0$. Each curve is labeled by the corresponding value of the plasma parameter. The dashed curves depict the solution of H17 for the same set of the input parameters. The solutions are qualitatively similar except close to the inner boundary since the H17 solution is valid for $x \gg x_{\rm in}$. Away from the inner boundary, differences between the solutions are small when $\beta_0$ is large. In Fig. \ref{fig:ww}, we show that the stress component associated with wind launching,  $W_{z\phi}$, is much smaller than the stress componenent due to disc turbulence $W_{r\phi}$ when the magnetic field strength is weak (i.e., large $\beta_0$). The term associated with the wind in equation (\ref{eq:steady-cw}), therefore, becomes smaller than the contribution corresponding to disc turbulence for large values of $\beta_0$. The greater the wind strength, the greater the difference between solutions. The surface density of H17 is smaller than the exact solutions (except close to the inner boundary) and for a case with $\beta_0 =10^3$, the density profiles differ by a factor of two.
In the bottom panel of Fig. \ref{fig:surf-a}, the surface density is shown for a larger temperature exponent, $q=1$. Again,the H17 surface density  is always lower than our solutions, irrespective of the wind strength. 
Thus, it can be concluded that the approximation used by H17   underestimates the surface  density, up to a factor of two when the wind is strong.

In Fig. \ref{fig:surf-b}, we display the radial surface density profile for different temperature exponent and plasma parameter. The mass of the central star is $M_{\star}=1$ M$_{\odot}$ and the accretion rate is assumed to be $\dot{M}_{\rm d}=10^{-8}$ M$_{\odot}$ yr$^{-1}$. The solid curves correspond to $q=1/2$, whereas the dashed curves represent cases with $q=1$. Each curve is marked by the corresponding value of $\beta_0$. Significant surface density reduction is seen as the wind gets stronger, irrespective of the temperature exponent. However, the surface density becomes steeper with decreasing the temperature exponent. For a case with $q=1$, however, the  inner parts of the disc exhibit  a  steep profile.

 In the next section, we consider the effect of wind mass-loss and obtain new analytical solutions. With these solutions, we demonstrate that  the disc surface density profile is modified dramatically as a result of mass removal by the winds.

\subsection{Solutions with mass-loss, $\xi_{ w}\neq 0$}
When the wind mass-loss is not negligible and the disc is in a steady-state, equation (\ref{eq:main}) can be written as 
\begin{equation}\label{eq:steady-neq}
\frac{d}{dx} \left [ \sqrt{x} \frac{d}{dx} \left ( \xi_{ r} x^{2-q} y \right ) + \xi_{ z} x^{(2-q)/2} y \right ] - \xi_{w} \frac{y}{\sqrt{x}}=0 .
\end{equation}
Interestingly, this equation can be transformed to the well-known Whittaker equation and its general solution for  $q\neq 1$ is written in terms of Whittaker functions \citep[e.g.,][]{abra} 
For $q=1/2$, the general solution is 
\begin{multline}
y(x)=x^{-9/8}  \left [ C_{1} {\rm M}_{\mu,1}(z) +C_{2} {\rm W}_{\mu, 1}(z) \right ]  \\
\times  \exp \left [ 2 \left ( \xi_z / \xi_r  \right ) x^{-1/4} \right ],
\end{multline}
where ${\rm M}_{\mu,1}(z)$ and ${\rm W}_{\mu,1}(z)$ are Whittaker functions. In addition, the parameter $\mu$ and the independent variable $z$ are defined as
\begin{equation}
\mu = -\frac{3}{2}\xi_z \left ( 4\xi_r \xi_w + \xi_{z}^2 \right )^{-1/2}, 
\end{equation}
and
\begin{equation}
z=4 \left [ 4 (\xi_w / \xi_r ) + \left( \xi_{z} / \xi_r \right)^2 \right ]^{1/2} x^{-1/4}.
\end{equation}

With $q=1$, the general solution of equation (\ref{eq:steady-neq}) is  a combination of two power-law functions
\begin{equation}
y(x)=C_1 x^{-n_1} + C_2 x^{-n_2},
\end{equation}
where the exponents $n_1$ and $n_2$ may be written as
\begin{equation}
n_{1,2} = A\pm B,
\end{equation}
with
\begin{equation}
A = \frac{1}{4}  [3+ 2\left( \xi_z / \xi_r \right)]
\end{equation}
and
\begin{equation}
B = \frac{1}{4} 
\left [1+16 \left( \xi_w / \xi_r \right)+4\left( \xi_z /  \xi_r \right)+4 \left( \xi_z / \xi_r \right)^2 \right ]^{1/2}.
\end{equation}

It is worth noticing that, for a given value of $q$, the constants of integration $C_1$ and $C_2$ are obtained by imposing proper boundary conditions. In particular, $C_1$, is obtained by demanding that equation (\ref{eq:boundary_I}) is satisfied, whereas $C_2$ is obtained from equation (\ref{outer-boundary-wind}) by providing the accretion rate at the outer disc edge.

\begin{figure*}
\includegraphics[scale=1.2]{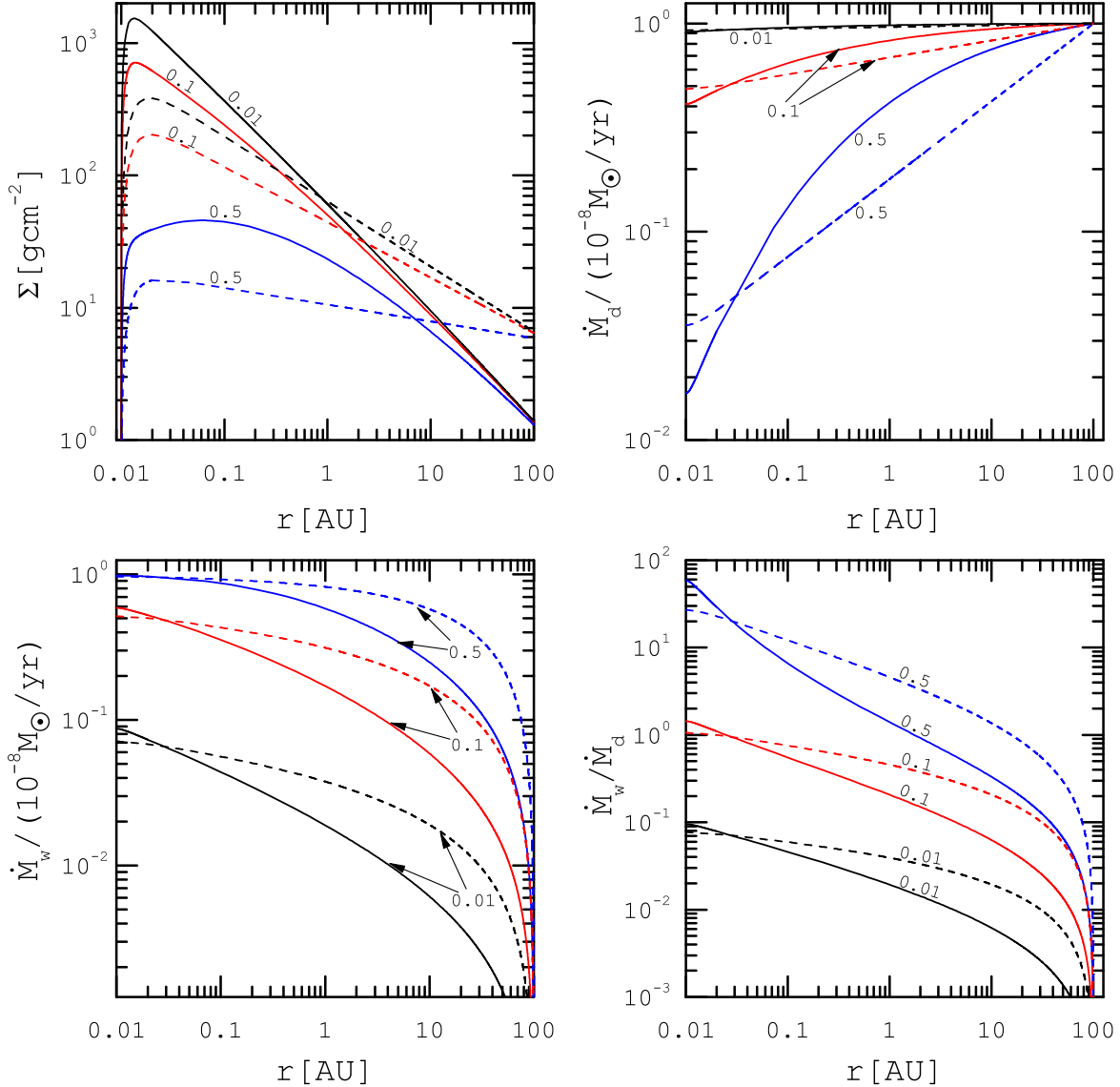}
\caption{
Radial disc structure obtained for $\beta_0 =10^4$ and temperature exponents $q=1/2$ (solid curves) and $q=1$ (dashed curves). Each curve is labeled by  the ratio $\psi=\xi_{w}/\xi_z =0.01$, $0.1$, and $0.5$ (black, red, and, blue lines respectively). The central mass considered is $M_{\star}=1$ M$_{\odot}$ and the accretion rate at the outer edge of the disc, i.e., $r_{\rm out}=100$ AU, is assumed to be $\dot{M}_{\rm d}=10^{-8}$ ${\rm M}_{\odot}/{\rm yr}$. On the top, left panel the disc surface density profile is shown.  On the top, right panel the normalized gas accretion rate, $\dot{M}_{\rm d}/(10^{-8} M_{\odot}{\rm yr}^{-1})$. Radial profile of the  mass loss rate by the wind, $\dot{M}_{\rm w}/(10^{-8} M_{\odot}{\rm yr}^{-1})$, is shown on the bottom, left panel and the ratio $\dot{M}_{\rm w}/\dot{M}_{\rm d}$ is presented on the bottom, right panel.
}
\label{fig:surf-mdot}
\end{figure*}

Although the parameters $\xi_r$ and $\xi_z$ characterizing the stress are expressed in terms of the plasma parameter $\beta_0$, we do not have a relation between the wind mass-loss parameter $\xi_w$ and the plasma parameter $\beta_0$. However, as more angular momentum is transported by the wind, the mass loss by the wind increases. This simplifying assumption  enables us to write $\xi_w$ directly proportional to $\xi_z$, i.e. $\xi_w = \psi \xi_z$, where the constant of proportionality $\psi$ is an input parameter of our model.

Fig. \ref{fig:surf-mdot} displays the radial profile of the disc quantities for  $\beta_0 =10^4$ and different values of parameter $\psi$. The solid curves represent solutions with $q=1/2$, and the dashed curves correspond to cases with $q=1$. Each curve is marked by the corresponding value of the mass-loss parameter $\psi$.  The inner and outer edges of the disc are at $r_{\rm in}=0.01$ AU and $r_{\rm out}=100$ AU. The accretion rate at the outer edge is assumed to be $\dot{M}_{\rm out}=10^{-8}$ M$_{\odot}$/yr. Furthermore, the central mass is $M_{\star}=1$ M$_{\odot}$. In the top left panel, the surface density is shown. Due to the wind mass-loss, not only does the disc surface density decrease, but also the radial slope of the surface density distribution is significantly changed. As the wind extracts more mass from the disc, the surface density becomes shallower, so that when the mass-loss is very efficient, the slope becomes zero and even has a positive gradient in a progressively larger region beyond the inner boundary.

This  trend has already been found in the time-dependent studies of a disc with a wind \citep[e.g.,][]{Suzuki16} and posses significant astrophysical implications. For instance, radial migration of a planet can be halted due to the surface density with a positive slope \citep[e.g.,][]{Ogihara2015}. Also, due to the fact that the radial motion of dust particles is strongly dependent on the radial gradient of the pressure, our solutions suggest that the dynamics of these particles in a disc with strong wind is very different from a case in which the wind is not present. Although exploring these topics is beyond the scope of the present study, our analytical solutions  provide a good framework for such studies. 

In the top right panel of Fig. \ref{fig:surf-mdot} 
the normalized accretion rate, $\dot{M}_{\rm d}/(10^{-8} {\rm M}_{\odot}/{\rm yr})$, versus the radial distance is shown. In the absence of a wind, it is reasonable to expect that the accretion rate will remain spatially constant. When wind mass-loss is considered, the closer we get to the inner region, the more the accretion rate is reduced. For instance, if we set $\psi=0.5$, the inner accretion rate is reduced by a factor of about 100 compared to the outer edge accretion rate. In the bottom left panel, the rate of mass-loss by the wind is shown. Wind-mass loss is negligible  in the outer region of the disc, however, this rate gradually becomes significant in the inner regions. The ratio of the wind mass-loss rate and the accretion rate as a function of the radial distance is shown in the bottom right panel. When the mass-loss parameter is $\psi =0.01$, the ratio is less than unity which means that most of the disc gas is accreted onto the central star rather being extracted by a wind. If a larger value is adopted for the mass-loss parameter, say, $\psi =0.5$, the trend is reversed and most of the accreting gas is lost by the wind. 

We note that our analytical solutions can be used to explore a larger parameter space, however, properties of the disc quantities are qualitatively similar to what we have found so far. In the next section, we discuss  astrophysical implications of the solutions, in particular regarding the HL Tau disc \citep{Kwon11,Yen2017}.

\section{Astrophysical Implications: HL Tau disc}

The model presented in this paper for PPD   structure with magnetically-driven winds and our analytical solutions  can be used to describe an accretion disc structure with winds. We now discuss how this model is used to describe certain observational feature of HL Tau disc. However,  the astrophysical implications of our solutions are not restricted just to those we mention here. 

According to the conventional viscous disc model, the  high accretion rate of HL Tau requires  a high level of turbulence and thereby, the dust layer is predicted to be thick, which is not consistent with the observations \citep{Kwon11,Yen2017}. 
H17 suggested that a magnetically-driven wind is able to extract the required angular momentum to maintain a high accretion rate, whereas the vertical scale-height of the dust layer is determined by dust diffusion. Their approach, however, involved a simplified disc structure with no wind mass-loss. We re-assess these findings using our new analytical solution that simultaneously accounts for wind-driven accretion and wind mass-loss.

As for the key observational features of HL Tau, H17 considered estimated disc properties based on either observations or comparisons between observations and radiative transfer modeling \citep{Kwon11,Akiyama,Car,Pinte16}. In order to ease the comparison with the results obtained with our model, we adopt their input parameters.  In particular, we adopt for the midplane disc temperature profile of HL Tau $T_{\rm mid}=62 (r/{\rm AU})^{-0.43}$K, as found in \cite{Kwon11}.

There are three main constraints imposed by either observations or physical conditions that guide the study that follows. 
\\
{\it i)} The dust layer in HL Tau is surprisingly thin, with a thickness of around 1 AU at the radial distance of 100 AU \citep{Pinte16}. According to H17, this vital observational feature of HL Tau can be explained if wind-driven accretion occurs in this system. \\
{\it ii)} The disc model we propose is no longer valid if the Toomre parameter, $Q=c_{\rm s}\Omega /\pi G \Sigma$, \citep{Toomre} drops to a values less than its threshold value $Q_{\rm c}\simeq 2$ \citep[e.g.,][]{Voro}. In the regions where the Toomre parameter is less than this critical value, the disc is subject to  gravitational instability and gravity-driven turbulence may play a dominant role in transporting angular momentum \citep[e.g.,][]{Rice2005}. Since this is not included in our model, or in H17, we can not determine consistently the structure of the disc in regions associated with $Q$ values beyond the stability threshold.   \\
{\it iii)} Dust growth calculations show that the maximum size of dust particles has a Stokes number of about 0.1 \citep[e.g.,][]{Brauer,Okuzumi}. For dust particles with radius $a<1-10$ mm in a PPD, the Stokes number is written as ${\rm St}=\pi \rho_{\rm s} a/2\Sigma$, where $\rho_{\rm s}=2$ g cm$^{-3}$ is the material density of a dust particle \citep[e.g.,][]{Youdin,Miyake2016}

Using their disc model with no wind mass-loss, H17 investigated under what circumstances the above three conditions are satisfied simultaneously in the presence of magnetic winds. They found that a very limited range of the input parameters is able to reproduce the configuration of the HL Tau disc. In what follows, we explore the impact of considering models that account for wind mass-loss and show that the observational constraints can be satisfied for a larger region of the parameter space defining the disc model with winds. 


The inner and outer disc edge radii of HL Tau disc are poorly constrained by current observations, however, in order to proceed further we assume that the inner and the outer radii of the disc are 0.01 AU and 1000 AU, respectively. Although it has been suggested that the outer edge is about 1000 AU \citep{Yen2017}, the size of HL Tau disc can be smaller \citep{Kwon11}. We find that properties of the disc weakly depend on the adopted inner and outer disc radii. The accretion rate at the outer edge is assumed to be $\dot{M}_{\rm out}=1\times 10^{-7}$ M$_{\odot}$ yr$^{-1}$ and the central mass is $M_{\star}=1\,$M$_{\odot}$ \citep{Pinte16}.  Here, we also consider these two extreme cases,
for high mass-loss rate ($\psi=0.5$) and low mass-loss rate ($\psi=10^{-3}$).
This enables us to explore the effect of wind mass-loss on the configuration of HL Tau disc.

Before discussing the conditions under which all the above mentioned observational  conditions are satisfied, we display the surface density profile for different sets of the input parameters to determine regions where they are gravitationally unstable. The general solution of equation (\ref{eq:steady-neq}) for $q=0.43$ can be written as
\begin{multline}
y(x)=x^{-\frac{471}{400}}  \left [ C_{1} {\rm M}_{\mu,\frac{50}{57}}(z) +C_{2} {\rm W}_{\mu, \frac{50}{57}}(z) \right ]  \\
\times  \exp \left [ \frac{100}{57} \left ( \xi_z / \xi_r  \right ) x^{-\frac{57}{200}} \right ].
\end{multline}
In addition the parameter $\mu$ and the independent variable $z$ are defined as
\begin{equation}
\mu = -\frac{157}{114} \xi_z ( 4\xi_r \xi_w + \xi_{z}^2 )^{-1/2}, 
\end{equation}
\begin{equation}
z=\frac{200}{57} \left [4\left(\xi_w / \xi_r \right) + \left(\xi_z / \xi_r \right)^2 \right ]^{1/2} x^{-\frac{57}{200}}.
\end{equation}

\begin{figure}
\includegraphics[scale=0.6]{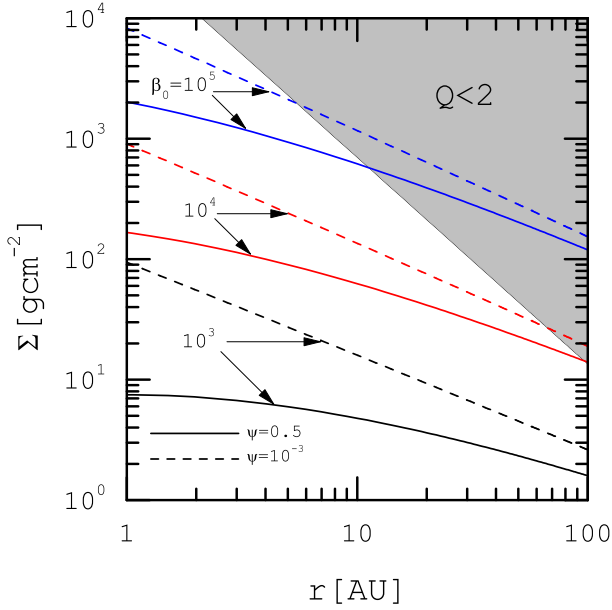}
\caption{Gas surface density distribution versus the radial distance for different values of the plasma parameter $\beta_0$ and two different rates of wind  mass loss, specified using the parameter $\psi$. The region with $Q<2$ is gravitationally unstable and is denoted by the shaded region. Each curve is labeled by the adopted value of $\beta_0$. with solid and dashed curves corresponding to the cases with $\psi=0.5$ and $10^{-3}$, respectively. 
}
\label{fig:den-Q}
\end{figure}

\begin{figure}
\includegraphics[scale=0.65]{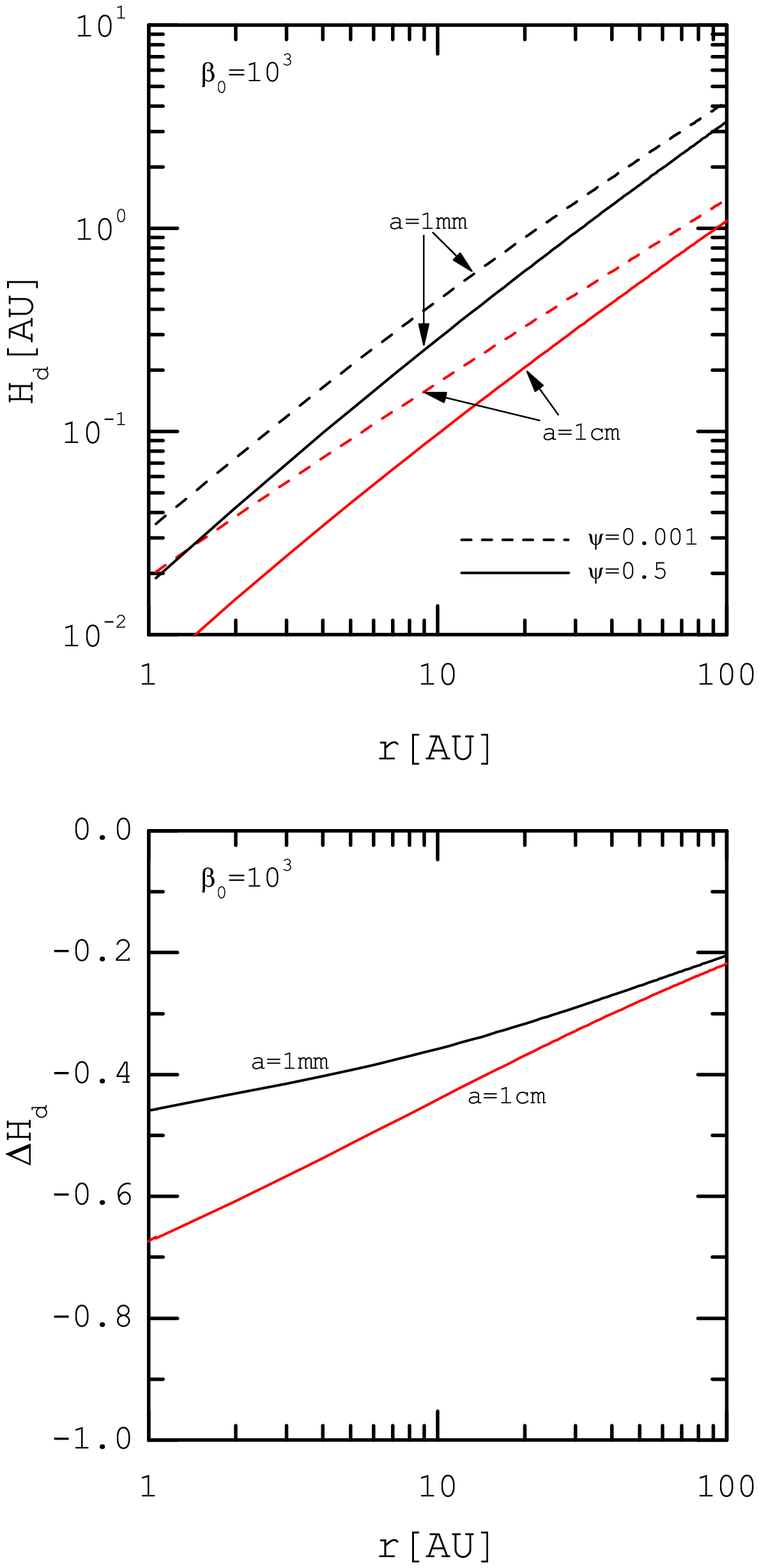}
\caption{Profile of the dust layer thickness, $H_{\rm d}$, versus the radial distance for a case with strong magnetic field ($\beta_0 =10^3$) and different particle size (top). Cases with $\psi =10^{-3}$ (i.e., low wind mass-loss rate) and $\psi=0.5$ (i.e., high wind mass-loss rate) are shown by solid and dashed curves, respectively. Each curve is labeled by the adopted particle size, i.e. $a=1$ mm and 1 cm. The bottom panel displays the radial profile of the ratio $\Delta H_{\rm d}=[H_{\rm d}(0.5)-H_{\rm d}(10^{-3})]/H_{\rm d}(10^{-3})$, where $H_{\rm d}(10^{-3})$ and $H_{\rm d}(0.5)$ denote dust layer thickness for $\psi=10^{-3}$ and 0.5, respectively, for a given dust size. }\label{fig:thickness}
\end{figure}

Fig. \ref{fig:den-Q} shows the surface density profile as a function of the distance for different values of the parameters $\beta_0$ and $\psi$. Each curve is marked by the adopted value of $\beta_0$, whereas cases with $\psi=0.5$ and $10^{-3}$ are shown by solid and dashed curves, respectively. As before, we find that as the wind strength increases (i.e., for higher $\psi$), it carries more mass and the surface density reduction becomes more significant, particularly in the inner region. The fate of the gravitational perturbations in a disc is quantified in terms of the Toomre parameter, which depends only on the surface density distribution for a disc with Keplerian rotation and a given temperature distribution. According to linear perturbation studies and numerical simulations of self-gravitating discs, once the Toomre parameter becomes less than a critical value, the disc becomes gravitationally unstable. Under these circumstances, gravitational instability may dominate angular momentum transport and furthermore lead to the formation of the gaseous clumps. Although there are considerable debates on the exact value of the Toomre threshold value and the efficiency of the cooling mechanisms in triggering gravitational instability, following previous studies we consider the Toomre threshold value is 2. 
We note that some authors take $Q=1$, rather than $Q=2$, as a critical value. The results obtained depend only weakly on the critical $Q$ value considered in this range. Adopting the latter value enables us to make a direct comparison with H17. 
The shaded region in Fig. \ref{fig:den-Q} denotes the region where instability criterion,  $Q<2$, is satisfied.  For relatively weak winds (i.e., $\beta_0 = 10^5$) the disc structure can be visualized as an inner gravitationally stable part and an outer gravitationally unstable region. The radial extent of the stable region, however, expands as $\beta_0$ decreases and more mass is removed by a stronger wind. Disc models with intermediate and strong winds are gravitational stable for $r\lesssim 100$\,AU irrespective of the wind mass-loss rate.

Since the second observational condition is a restriction on the dust layer thickness $H_{\rm d}$,  the radial profile of $H_{\rm d}$ is shown in  Fig. \ref{fig:thickness} for particles with radii $a=1$ mm and $1$ cm and different values of $\psi$ . The vertical distribution of the dust layer is determined by the dust diffusion and dust settling due to the drag force. Therefore, the dust layer thickness is written as \citep{Youdin},
\begin{equation}\label{eq:Hd}
H_{\rm d} = \left (1+\frac{\rm St}{\alpha_{D}} \right )^{-1/2} H,
\end{equation}
where the gas component thickness is $H=\sqrt{2} c_{\rm s}/\Omega$, and, $\alpha_{\rm D}$ is the normalized diffusion coefficient: $\alpha_{\rm D}=D_z / c_{\rm s} H$. Here, $D_z$ denotes the vertical diffusion coefficient for small particles. With numerical simulations, the normalized diffusion coefficient $\alpha_{\rm D}$ can be expressed in terms of the parameter $\beta_0$ by the following fitted relation \citep{Zhu15}:
\begin{equation}\label{eq:alphaD}
{\rm log} \alpha_{\rm D} = 1.1 - {\rm log}\beta_{0}.
\end{equation}

\begin{figure}
\includegraphics[scale=0.65]{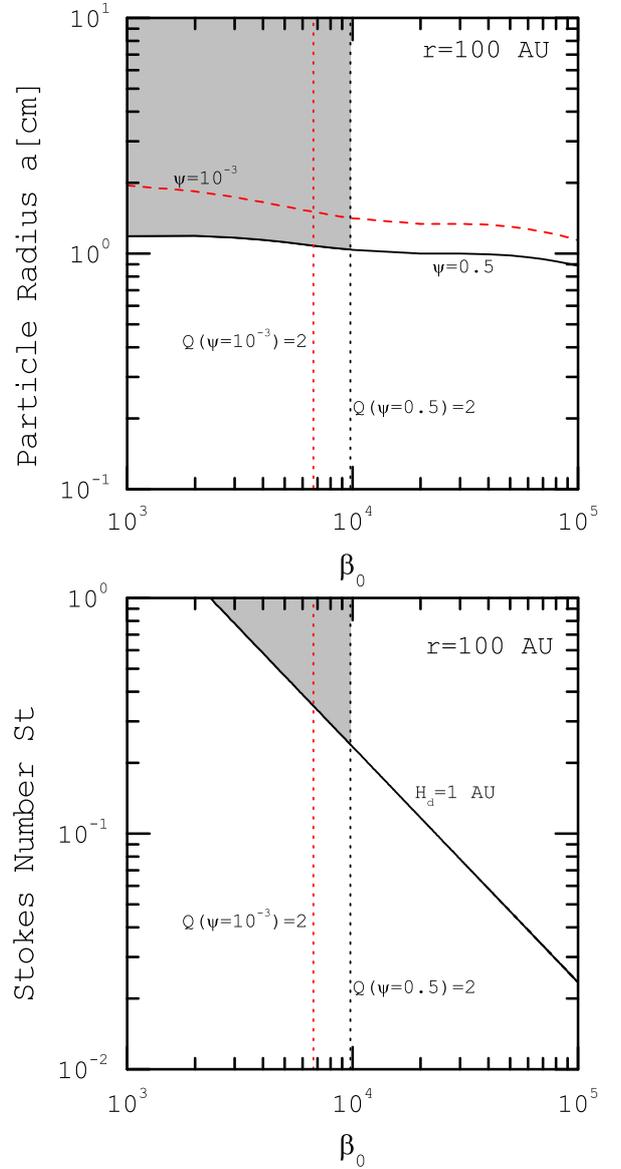}
\caption{Top panel shows radius of dust particles with $H_{\rm d}=1$ AU  at the distance 100 AU as a function of the plasma parameter $\beta_0$. Each curve is labeled by the adopted value of $\psi$. Vertical dotted lines represent the values of $\beta_0$ at which the Toomre parameter is equal to its threshold value 2. In the bottom panel, Stokes number St as a function of $\beta_0$ is shown for the particles with thickness 1 AU at the distance 100 AU. }\label{fig:r100AU}
\end{figure}

\begin{figure}
\includegraphics[scale=0.65]{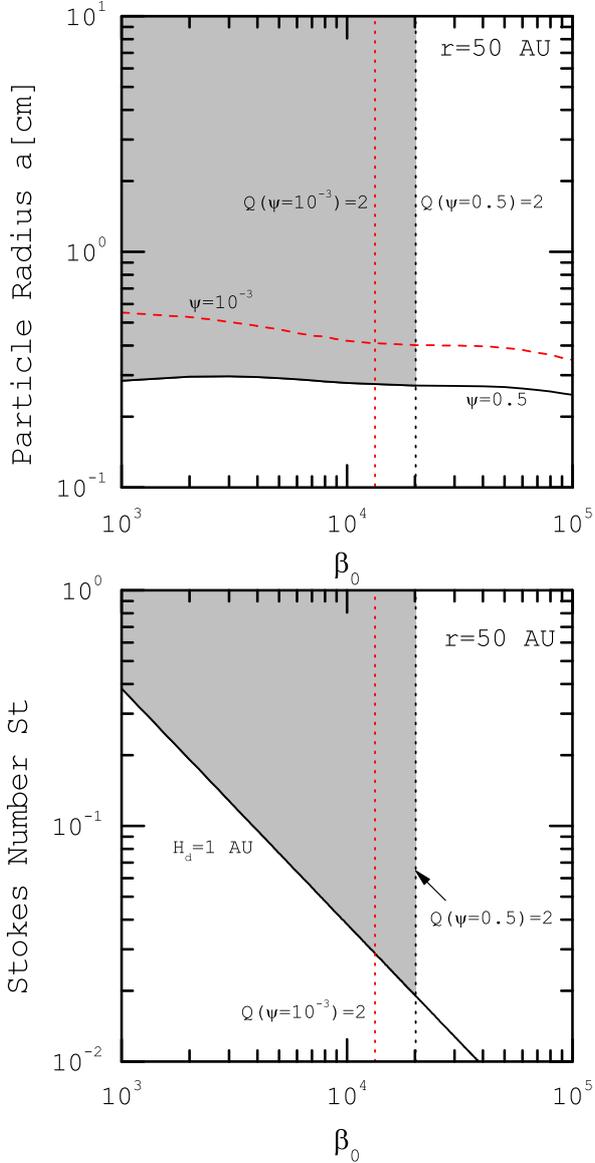}
\caption{Similar to Fig. \ref{fig:r100AU}, but at the radial distance 50 AU.}\label{fig:r50AU}
\end{figure}

Fig. \ref{fig:thickness} (top) shows the dust layer thickness $H_{\rm d}$ as a function of the radial distance for particles with sizes 1 mm and 1 cm and different values of the mass-loss parameter $\psi$. The input parameters are the same as the parameters used in Fig. \ref{fig:den-Q}, however, a strong magnetic field configuration with $\beta_0 = 10^3$ is considered. The solid curves correspond to a case with significant wind mass-loss rate (i.e., $\psi =0.5$), whereas the dashed curves belong to a case where wind does not carry considerable mass (i.e., $\psi =10^{-3}$). Furthermore, each curve is marked by the adopted particle size. This figure shows that the dust layer thickness is reduced when wind mass-loss is considered and the effect is more significant for larger particles. The effect of mass removal by the wind is also shown in the bottom panel of Fig. \ref{fig:thickness}. This panel shows radial profile of the parameter $\Delta H_{\rm d}$ which is defined as 
\begin{equation}
\Delta H_{\rm d} = \frac{H_{\rm d}(\psi=0.5) - H_{\rm d}(\psi=10^{-3})}{H_{\rm d}(\psi=10^{-3})},
\end{equation}
where $H_{\rm d}(\psi=0.5)$ and $H_{\rm d}(\psi=10^{-3})$ denote the dust thickness for $\psi=0.5$ and $10^{-3}$, respectively, and a given dust size. This shows that dust layer thickness reduces significantly in the presence of winds. Our analysis, therefore, supports the main suggestion of H17 for explaining the HL Tau disc based on a wind-driven disc model. We have shown that when wind mass-loss is included, the dust layer is thinner and that is exactly a key advantage of the present model comparing to  a conventional viscous disc model.

We now explore the parameter space for which the particle size and the Stokes number fulfill the required observational constraints. Fig. \ref{fig:r100AU} depicts  results for a wide range of $\beta_0$ and a given dust layer thickness, $H_{\rm d}=1$ AU, at the radial distance $100$ AU. As before, cases with high (i.e., $\psi=0.5$) and low (i.e., $\psi=10^{-3}$) mass-loss rates are considered and each curve is labeled by the mass-loss parameter $\psi$. The vertical dotted lines specify the value of $\beta_0$ at which the Toomre parameter becomes 2 at 100 AU. This critical value of $\beta_0$ shifts to a larger value as more mass is lost by the wind. In the top, the particle size with $H_{\rm d}=1$ AU and for $\psi=0.5$ is shown by the solid black curves. 
The area between these curves which satisfies both constraints $H_{\rm d} \lesssim 1$ AU and $Q \gtrsim 2$ is shown as a shaded region. Similar curves are also shown for a case with low mass-loss rate $\psi=10^{-3}$ as a red dashed curve for the particle size and a red vertical dotted line for the Toomre parameter. When the mass-loss rate is low, the size of the acceptable shaded region shrinks. We note, however, that the allowed region for low mass-loss rate is slightly smaller than H17. There are three main reasons to explain this discrepancy. First of all, H17 analysis is based on an approximate surface density profile where, as we showed earlier, their solution underestimates the gas surface density. Furthermore,  the approach employed by H17 does not include the effect of wind-driven mass-loss, whereas we allowed the disc to lose part of its accreting mass via a wind, even for small values of the parameter $\psi$. Finally, we note that the disc scale height is frequently defined as $H=c_{\rm s}/\Omega$, instead of our implemented equation, i.e. $H=\sqrt{2}c_{\rm s}/\Omega$. H17 adopt the former, however, our relation for the disc scale height includes a numerical factor $\sqrt{2}$.

In Fig. \ref{fig:r50AU}, we repeat the analysis of Fig. \ref{fig:r100AU}, but at a radial distance $50\,\rm AU$. Observational knowledge of the HL Tau dust layer thickness is poorly constrained at a distance smaller than $100\,\rm  AU$. As an illustrative case, we also consider an extreme situation in which dust thickness at $50 \,$AU is same as dust thickness as at $100 \,$AU, i.e. $H_{\rm d}=1\,$AU.  Fig. \ref{fig:r50AU} displays similar trends as in Fig. \ref{fig:r100AU}. The wind mass-loss expands the  allowed region of the parameter space, though if a smaller fixed dust layer was adopted, the accepted region would be more restricted.

\section{Conclusions}
To describe the structure of a PPD with a magnetically-driven wind, we have presented a simplified model within the framework of the standard accretion disc model, however,  angular momentum transport by the disc turbulence and a wind and mass-loss rate  are parameterized in terms of plasma parameter $\beta_0$ using disc MHD simulations. In this regard, our work differs from previous (semi)analytical works where these input parameters are adopted independently. Since numerical simulations of discs with MRI-driven winds are computationally expensive, the physical parameter space that one can explore is limited by this restriction. Therefore, an analytical disc wind solution, that we obtained here, offers the advantage of studying an accretion disc with MRI-driven winds over a wide range of the input parameters. 

A similar model has been implemented by H17 to explain the high accretion rate  of HL Tau and its  efficient dust settling. However, the analysis of H17  is based on an approximate solution that includes  no mass-loss. By relaxing their assumptions, we  obtained analytical solutions in a case without wind mass-loss and found that H17 solution underestimates the surface density in particular when the magnetic field is strong. Our model incorporates the disc wind finds a wider range of acceptable parameter space to explain the observational features of HL Tau. 

Although our focus is to explore disc structure with MRI-driven winds, we should note that a PPD is subject to the radiation of its central star or even external radiation \citep{alex2006,alex2006b,adams2004,gorti2009}. Under these circumstance a photoevaporative wind may be able to create an inner hole and even efficiently disperse the entire disc \citep{alex2006b,gorti2009}.  MRI-driven  winds are also able to reduce the disc surface density and  remove a significant fraction of the disc angular momentum and trigger rapid accretion \citep{Armitage13}. It is not clear yet which of these mechanisms dominate in a PPD, however, an interesting direction for future work is to explore observational consequences of either processes.  \cite{Carrera17}, for instance, investigated planetesimal formation in a disc subject to photoevaporative winds and we propose a similar study  in a disc with MRI-driven winds using disc wind solutions obtained here. 

\cite{Suzuki2010} showed that in PPDs an inside-out clearing can occur due to presence of disc winds. With numerical simulations, they constrained a set of the input parameters for exploring evolution of a disc within the framework of the standard disc model. A direct comparison between their analysis and our model is not  appropriate, because our calculations are based on the steady state approximation, whereas they explored evolution of a disc with winds. Furthermore, they did not consider angular momentum transport by MRI-driven disc winds. Our solutions, however, demonstrate that the surface density reduction due to the disc winds is more significant in the inner region comparing to the outer part of a disc and this reduction is more evident as the wind becomes stronger and more mass is extracted by the wind. Perhaps if time-dependent calculations were made, we could recover the inside-out clearing of PPDs due to a magnetic disc wind as \cite{Suzuki2010} found  in their study. 

The dynamics of dust particles in PPDs is an important topic, for it provides the essential key information on how these particles are accumulated and eventually form planets. Since the drag force and the gravitational force of the central star determine the net force on a dust particle, its delivery and possible growth rate significantly depends on the disc structure. When the associated Stokes number of a particle becomes of order unity, particles are subject to radial migration towards the central star \citep{Lamb12,Lamb14}. These so-called pebbles, however, can be captured by the gravitational field  of an  already existing kilometer-sized planetesimal and accreted onto it via drag-driven process \citep{Ormel2010}. The subjects of particle growth to pebble-size and planet formation via pebble accretion have attracted considerable attention over recent years, though as far as we know, most of the related studies are restricted to a PPD without winds \citep{Lamb12,Lamb14}. Recently, \cite{shadmehri17} investigated pebble delivery in a PPD with MRI-driven winds using the H17 approximate solution. They showed that in a wind-driven accretion disc growth rate of the particles to the pebble-sized and their delivery are reduced in the presence of winds. A missing key process in \cite{shadmehri17} is the wind mass-loss. It would be interesting to examine the dynamics of pebbles in a PPD with magnetic winds using the present analytical solutions in which both angular momentum and mass-loss by winds are considered.

\section*{Acknowledgements}
MS and FK thank the hospitality and support during their visit to the Niels Bohr International Academy at The Niels Bohr Institute where part of this work was carried out. This work has been supported financially by Research Institute for Astronomy \& Astrophysics of Maragha (RIAAM) under research project No. 1/5440-7 (FK). 
The research leading to these results has received funding from the European Research Council (ERC) under the European Union's Seventh Framework programme (FP/2007-2013) under ERC grant agreement No 306614 (MEP).
RGM acknowledges support from NASA through grant NNX17AB96G.

\bibliographystyle{mnras}
\bibliography{reference} 




\bsp	
\label{lastpage}
\end{document}